\documentclass[12pt]{article}

\newcommand{\rmi}{{\rm i}}
\newcommand{\rme}{{\rm e}}
\newcommand{\rmd}{{\rm d}}
\newcommand{\klong}{|K_{\rm L}\rangle}
\newcommand{\kshort}{|K_{\rm S}\rangle}
\newcommand{\knot}{|K^0\rangle}
\newcommand{\knotbar}{|\bar{K}^0\rangle}
\newcommand{\tf}{t_{\mbox{\tiny flight}}}

\usepackage{amsthm,graphicx,cite}
\usepackage{epsf,latexsym}
\usepackage{amssymb,amsmath}

\begin{document}

\def\llra{\relbar\joinrel\longrightarrow}              
\def\mapright#1{\smash{\mathop{\llra}\limits_{#1}}}    
\def\mapup#1{\smash{\mathop{\llra}\limits^{#1}}}     
\def\mapupdown#1#2{\smash{\mathop{\llra}\limits^{#1}_{#2}}} 

\title{Time as a dynamical variable in quantum decay}

\author{Rafael de la Madrid \\
\small{\it Department of Physics, Lamar University,
Beaumont, TX 77710} \\
\small{E-mail: \texttt{rafael.delamadrid@lamar.edu}}}

\date{\small{June 21, 2013}}









\maketitle

\begin{abstract}
\noindent We present a theoretical analysis of quantum decay in which the 
survival probability is replaced by a decay rate that is equal to 
the absolute value squared of the wave function in the time 
representation. The wave function in the time representation
is simply the Fourier transform of the wave function in the 
energy representation, and it is also the probability amplitude generated
by the Positive Operator Valued Measure of a time operator. The present 
analysis endows time with a dynamical character in quantum decay, 
and it is applicable only when the unstable system is 
monitored continuously while it decays. When 
the analysis is applied to the Gamow state, one recovers the exponential
decay law. The analysis allows us to interpret the
oscillations in the decay rate of the GSI anomaly,
of neutral mesons, and of fluorescence quantum beats as the result
of the interference of two resonances in the time representation. In 
addition, the analysis allows us to show that the time of flight of a
resonance coincides with its lifetime.
\end{abstract}




\section{Introduction}
\setcounter{equation}{0}
\label{sec:introduction}

In quantum mechanics, time plays the role of an external parameter,
and therefore it is apparently not a dynamical variable, as made clear
by Pauli's theorem~\cite{PAULI}. However, there are many experimental 
situations such as 
the time of flight or the decay of an unstable particle in which 
time seems to play a dynamical role. For example,
the lifetime of a particle seems to be an intrinsic dynamical property of the 
particle, not just a mere parameter. 

Many authors have constructed time operators that endow time with a 
dynamical character, see for 
example Refs.~\cite{HOLEVO,MUGA1,MUGA2,BUSCH,SRINIVAS,GIANNITRAPANI,
RECAMI,HEGERFELDT,MOYER,GALAPON,JULVE,EGUS,ROVELLI,HATANO} and references therein. Such time operators are 
usually~\cite{HOLEVO,MUGA1,MUGA2,BUSCH,SRINIVAS,GIANNITRAPANI,JULVE,EGUS,HEGERFELDT,MOYER} 
associated with Positive
Operator Valued Measures (POVMs) and therefore circumvent Pauli's theorem.
POVMs not only provide a natural setting for time operators, but also
for phase operators and for the momentum operator of a one-dimensional particle
on the half line. Rather than
being uncommon, POVMs are
standard tools in the quantum theory of open systems~\cite{DAVIES} and 
in quantum information and computation~\cite{PRESKILL,NIELSEN}.

Although the mathematical aspects of the POVMs associated with time operators 
are well established, their phenomenological signatures have remained
elusive~\cite{MIELNIK}. The 
purpose of this paper is to propose a theoretical analysis of quantum decay 
in which the decay rate is given by the probability distribution 
associated with the POVM of a time operator. In such analysis,
time appears explicitly as a dynamical variable (or, more 
precisely, as a random variable). We will show that
the probability distribution associated with the POVM of the time operator 
is different from the survival probability. We will 
also show that the time representation of the Gamow states describes 
the exponential region of quantum decay while explicitly 
displaying the dynamical character of time. 

As we will stress along the paper, describing the decay on an unstable
system in the time representation is necessary only in experiments that
monitor the system's decay continuously. One such experiment is the
so-called GSI anomaly~\cite{GSI}, where Litvinov {\it et al.}~observed 
that K-shell electron capture decay rates of 
Hydrogen-like $^{140}{\rm Pr}^{58+}$
and $^{142}{\rm Pm}^{60+}$ ions show an oscillatory modulation superimposed 
on the exponential decay. Because 
Litvinov~{\it et al.} monitored individual ions continuously, we will
interpret the GSI anomaly as the result of
the interference of two resonances in the time 
representation. We will also see that such interpretation could be applied
to the decay of $K$ and $B$ mesons and to fluorescence
quantum beats if the decay of these systems were monitored 
continuously.

In Sec.~\ref{sec:radecay}, we recall the basic phenomenological features
of exponential decay. In Sec.~\ref{sec:thdorte}, we recall the
standard theoretical analysis of quantum decay. In Sec.~\ref{sec:timerpe},
we construct the time representation and use dimensional analysis 
to identify the decay rate with the absolute value squared of the wave function
in the time representation. In Secs.~\ref{sec:timerepGS} 
and~\ref{sec:single-resys}, we obtain the time representation of a Gamow
state and show that such time representation accounts for the phenomenology
of exponential decay. In Sec.~\ref{sec:survvsdecay}, we compare the
survival probability $p_{\rm s}(\tau )$ with the non-decay probability 
${\cal P}(t)$ associated with the time representation,
and we point out that ${\cal P}(t)$ does not exhibit the Zeno effect. In 
Secs.~\ref{sec:gsianom} and~\ref{sec:kaon}, we 
show that the interference of two resonances in the time representation 
can account for the GSI anomaly, for fluorescence quantum beats, and for 
the decay of neutral mesons. In Sec.~\ref{sec:continuousme}, 
we compare the pulsed and the continuous measurements of the survival
probability $p_{\rm s}(\tau)$ with the measurement of the non-decay 
probability ${\cal P}(t)$, and we argue that the measurement
of ${\cal P}(t)$ is inherently continuous. In Sec.~\ref{sec:timeofflig}, 
we use the time representation to derive an expression for the
time of flight of a particle, and we show that the time
of flight of a resonance is equal to its lifetime, as it is usually 
assumed.
Section~\ref{sec:conclusions} contains our conclusions.

\section{Phenomenology of radioactive decay}
\setcounter{equation}{0}
\label{sec:radecay}

The standard phenomenological treatment of the decay of a 
radioactive sample is as follows. When a sample of radioactive nuclei 
contains $N(t)$ radioactive nuclei at time $t$, the rate at which nuclei decay 
is proportional to $N(t)$,
\begin{equation}
            \frac{\rmd N(t)}{\rmd t}=-\lambda N(t) \, ,
           \label{expndec}
\end{equation}
where $\lambda$ is the decay constant. Straightforward integration yields
\begin{equation}
         N(t) = N_0 \rme ^{-\lambda t} \, ,
          \label{numofprtime}
\end{equation}
where $N_0$ is the number of radioactive nuclei at $t=0$. The non-decay and the
decay probabilities are
\begin{eqnarray}
       && {\cal P} (t)=\frac{N(t)}{N_0}= \rme ^{-\lambda t} \, ,
                 \label{ndpN}  \\
       && {\cal P}_{\rm d}(t)= \frac{N_{\rm d}(t)}{N_0}= \frac{N_0-N(t)}{N_0} 
               =1-{\cal P}(t)=1-\rme ^{-\lambda t} \, , 
              \label{dpN}     
\end{eqnarray}
where $N_{\rm d}(t)$ is the number of atoms that have decayed at time $t$, 
that is, the number of detector clicks that result from observing 
the decay products of a radioactive reaction.

Quite often, as for example in Ref.~\cite{GSI}, we are interested in the 
decay rate. The decay rate is defined as 
\begin{equation}
    R(t)\equiv \frac{\rmd N_{\rm d}(t)}{\rmd t}= -\frac{\rmd N(t)}{\rmd t} \, ,
             \label{decminusin} 
\end{equation}
where the minus sign in Eq.~(\ref{decminusin}) comes from the fact that
the rate at which the mother nuclei have decayed is the
opposite to the rate at which such nuclei have {\it not} decayed. The decay 
rate also follows the exponential law,
\begin{equation}
               R(t)=R_0\rme ^{-\lambda t} \, ,
        \label{decayrate}
\end{equation}
where $R_0=\lambda N_0$. The decay rate has dimensions of 
probability/time (i.e., counts/time):
\begin{equation}
               [R(t)]=\frac{1}{\rm T} \, .
           \label{unitsdr}
\end{equation}
When we measure $R(t)$, we can obtain $N(t)$ from $R(t)$ by integration:
\begin{equation}
       N(t)= N_0-\int _0^t R(t^{\prime}) \rmd t^{\prime} \, .
         \label{numberofpar}
\end{equation}

When we measure the decay of a single radioactive nucleus (as it is done
in Ref.~\cite{GSI}), we need to repeat the experiment $N_0$ times, and the
above analysis carries through, except that the number of
initial radioactive nuclei $N_0$ is replaced by the number of times that 
we repeat the experiment. 

The output data of a decay experiment are usually expressed by plotting either
the number of decaying events (i.e., the number of detector ``clicks'') 
as a function of time, or the decay rate as a function of time. When
the system is monitored continuously, such output data can also be viewed 
as a temporal probability distribution of decay events, in very
much the same way that the output data of experiments that measure 
quantities such as arrival times, times of flight or tunneling times can
be viewed as temporal probability distributions of arrival, flight or tunneling 
events.

Because an unstable quantum system decays at a random time,
the measurement of quantum decay requires that we monitor
the system continuously, or else we may miss the moment when it decays.

\section{The standard theoretical treatment of quantum decay}
\setcounter{equation}{0}
\label{sec:thdorte}

Quantum mechanics describes the evolution of a system through
wave functions $\varphi (x;\tau )$ that
satisfy the time-dependent Schr\"odinger equation
\begin{equation}
         \rmi \hbar \frac{\rmd \varphi (x;\tau )}{\rmd \tau} =
             H\varphi (x;\tau ) \, ,
\end{equation}
where $\tau$ is a time {\it parameter} that labels the evolution of the
system and has no dynamical character~\cite{FN1}. In the position 
representation, the position operator acts as multiplication
by $x$, and Born's rule says that the probability density to find a particle at 
position $x$ is $|\varphi (x)|^2$. When the wave functions are normalized to 1,
$\int \rmd x \, |\varphi (x)|^2 =1$,
both $|\varphi (x)|^2$ and $|\varphi (x;\tau)|^2$ have 
dimensions of 1/length. In general, any given operator 
$A$ acts as multiplication by $a$ in the
$a$-representation (where $a$ runs over the
spectrum of $A$), and for any normalized wave function $\varphi$, 
$|\varphi (a)|^2$ has dimensions of 1/[$a$]. By Born's rule, $|\varphi (a)|^2$
is interpreted as a probability density.

It is customary to assume that the number of unstable particles that 
have not decayed at time $\tau$ is given by 
$N(\tau )=|\varphi (\tau)|^2$. The probability that the particle has not
decayed is then given by
\begin{equation}
       P(\tau )=\frac{N(\tau)}{N(0)}=
          \frac{|\varphi (\tau )|^2}{|\varphi (0)|^2} \, .
       \label{probstand}
\end{equation}
For a Gamow state of width $\Gamma _{\rm R}$, it can be easily shown that
\begin{equation}
       P(\tau )=\rme ^{-\Gamma _{\rm R}\tau /\hbar} \, .
           \label{ndamstand}
\end{equation}
However, because definition~(\ref{probstand}) assumes that time is just a 
parameter, we are going to construct a wave-function description of quantum
decay that utilizes the time representation.

\section{The Time Representation}
\setcounter{equation}{0}
\label{sec:timerpe}

In the remainder of this paper, we are going to work with a Hamiltonian
$H=H_0+V$, where $H_0$ is the free Hamiltonian and $V$ is a smooth, 
spherically symmetric potential. We will
assume that $V(r)$ is not too singular
at the origin and that it falls at infinity faster than exponentials 
(Appendix~A of Ref.~\cite{NPA08} contains the detailed mathematical 
characterization of the class of potentials we will use). We will
restrict ourselves to the s partial wave, since the generalization to
higher-order waves is straightforward. We will also restrict ourselves
to the continuous part of the spectrum, which will be assumed to be
$[0,\infty)$.

In order to construct the time representation, we first need to construct
the energy representation. We will use the 
energy representation associated with the ``out'' Lippmann-Schwinger 
eigenfunctions $\chi ^-(r;E)=\langle r|E^-\rangle$. The ``out'' 
energy representation of a wave function $\varphi (r)$ is given 
by~\cite{EXPLAMSI} 
\begin{equation}
       \varphi (E) = 
       \int_0^{\infty} \rmd r \, \varphi (r) \overline{\chi ^-(r;E)} \, , 
\end{equation}
which in Dirac's bra-ket notation reads
\begin{equation}
      \langle ^-E| \varphi \rangle = \int_0^{\infty} \rmd r \, 
           \langle ^-E|r \rangle \langle r| \varphi \rangle \, .
\end{equation}
In the energy representation, the Hamiltonian $H$ acts as multiplication by
$E$. Thus, by analogy to the position representation, where the 
position operator $Q$ acts as multiplication by $x$ and the momentum
operator acts as $P=-\rmi \hbar \rmd /\rmd x$, the time operator is 
usually defined as
\begin{equation}
         T\varphi (E) = -\rmi \hbar \frac{\rmd \varphi (E)}{\rmd E} \, .
             \label{deftimeop}
\end{equation}
Clearly, $T$ and $H$ satisfy the Heisenberg commutation relation
$[T,H]=-\rmi \hbar I$. However, although $T$ is a Hermitian operator (more
precisely, $T$ is a symmetric operator), it is not self-adjoint, and therefore
$T$ does not contradict Pauli's theorem. In particular,
the Hilbert-space spectrum of $T$ is the whole lower half of
the complex plane, including the whole real line. Because its 
Hilbert-space spectrum is complex, it seems that 
$T$ should be discarded. However, it has been realized that 
such operators can be justified if they are understood as 
POVMs~\cite{HOLEVO,MUGA1,MUGA2,BUSCH,SRINIVAS,GIANNITRAPANI,JULVE,EGUS,HEGERFELDT}. 


In quantum mechanics, POVMs usually arise whenever the system is embedded in a 
bath, and we trace out the bath degrees of freedom~\cite{DAVIES}. By
contrast, in the case 
of the time operator, the POVM arises because there is a lower bound in the 
energy, not because we are assuming that the system is embedded in a
bath. From the point of view of the quantum theory of measurement,
the POVMs of time operators arise because we are performing a continuous
measurement on the system~\cite{SRINIVAS}. 


For real values of $t$, i.e., for values of $t$ that have zero
imaginary part, the normalized eigenfunctions of $T$,
$\langle ^-E|t \rangle = \frac{1}{\sqrt{2\pi \hbar}}\rme^{\rmi Et/\hbar}$, 
are not delta-normalized 
but rather they satisfy 
\begin{equation}
     \langle t^{\prime}|t \rangle= 
        \int_0^{\infty} \rmd E \, \langle t^{\prime}| E^-\rangle 
            \langle ^-E|t \rangle =
           \frac{1}{2} \delta (t-t^{\prime}) +
            \frac{\rmi}{2\pi} {\rm P} \frac{1}{t-t^{\prime}} \, .
              \label{nonorth}
\end{equation}
From a calculational point of view, the non-orthogonality of the eigenfunctions
of $T$ for real $t$ is the main difference with respect to the case of 
a self-adjoint operator. Everything else, including the use of Dirac's
bra-ket notation, is very similar.

We can use the eigenfunctions
$\langle ^-E|t \rangle = \frac{1}{\sqrt{2\pi \hbar}}\rme^{\rmi Et/\hbar}$ 
to construct the time representation of a wave 
function by way of the Fourier transform~\cite{EXPLAMSI},
\begin{equation}
       \varphi (t) = \int_0^{\infty} \rmd E \, \varphi (E) 
        \frac{1}{\sqrt{2\pi \hbar}}\rme ^{-\rmi Et/\hbar} \, , 
            \label{timerewf}
\end{equation}
which in Dirac's bra-ket notation reads
\begin{equation}
      \langle t| \varphi \rangle = \int_0^{\infty} \rmd E \, 
           \langle t|E^- \rangle \langle ^-E| \varphi \rangle \, .
\end{equation}
Because for real $t$ the eigenfunctions of $T$ form a resolution 
of the identity,
\begin{equation}
      \int \rmd t \, |t\rangle \langle t|=I \, , 
\end{equation}
the probability density of a wave function is normalized to 1, as 
it should be,
\begin{equation}
       \int \rmd t \, |\varphi (t)|^2 =1 \, . 
\end{equation}
It is important to realize that $\varphi (t)$ is not the same as the
(unmonitored) time evolved state 
$\varphi (\tau )=\rme ^{-\rmi H\tau /\hbar}\varphi$~\cite{EXPLATI}. It 
is also important to realize that,
although the Hilbert-space spectrum of $T$ contains complex numbers, 
{\it only} real values of $t$ are used to calculate probabilities. Thus, 
similar to the case of a Hermitian Hamiltonian that produces 
resonances~\cite{CINVES}, the time operator's Hilbert-space 
spectrum (which is the lower half of the complex plane) does not coincide 
with its physical spectrum (which is just the real line).

In the time representation, the operator~(\ref{deftimeop}) acts as 
multiplication by $t$ (at least when it acts on wave functions
that satisfy $\varphi (E=0)=0$). Because any 
operator $A$ acts as multiplication by $a$ in the $a$-representation 
and $|\varphi (a)|^2$ is 
interpreted as the probability density that the measurement of $A$
(or, equivalently, the measurement of $|a\rangle \langle a|$) on
the state $\varphi$ yields the value $a$, we can 
interpret the operator~(\ref{deftimeop}) as the time operator and 
$|\varphi (t)|^2$ as the probability density 
that the measurement of $T$ 
(or, equivalently, the measurement of $|t\rangle \langle t|$)
on the state $\varphi$ yields the 
value $t$. Thus, the time representation is just like any representation 
associated with a self-adjoint operator, except that in the 
time representation one deals with POVMs
(rather than with projective measurements) and with probability densities
$|\varphi (t)|^2$ that represent temporal probability distributions of events.

Because $|\varphi (t)|^2$ has dimensions of 1/time,
Eq.~(\ref{unitsdr}) suggests that when $\varphi$ describes
an unstable state, it is not $N(t)=|\varphi (t)|^2$ but rather
\begin{equation}
   R(t) \equiv  |\varphi (t)|^2  \qquad 
             \mbox{(for one unstable particle)}\, ;
            \label{dr1}
\end{equation}
that is, the decay rate of a {\it single} unstable particle 
is given by the absolute value
squared of its wave function in the time
representation. If initially we have $N_0$ unstable particles, the decay rate is
\begin{equation}
    R(t) \equiv N_0|\varphi (t)|^2  \qquad 
             \mbox{(for $N_0$ unstable particles)} \, .
            \label{drn}
\end{equation}
Equations~(\ref{dr1}) and (\ref{drn}) are simply a rule to calculate 
probabilities in the time representation. Such rule is essentially the
same as the Born rule that we use to calculate probabilities 
in the representation 
associated with a self-adjoint operator. In fact, by combining
Eqs.~(\ref{ndpN}), (\ref{dpN}), (\ref{decminusin}) and (\ref{drn}), we can
write the decay and non-decay probabilities in terms of the wave function
in the time representation as follows:
\begin{equation}
       \frac{\rmd {\cal P}_{\rm d}(t)}{\rmd t} = |\varphi (t)|^2 
               \qquad 
             \mbox{(for one particle)}\, ;
        \label{dendindd}
\end{equation}
\begin{equation}
       \frac{\rmd {\cal P}(t)}{\rmd t} = -|\varphi (t)|^2
            \qquad 
             \mbox{(for one particle)}\, .
          \label{dendindd2}
\end{equation}
In the remainder of this paper, we will apply the
rule~(\ref{dr1})-(\ref{dendindd2}) to exponential decay,
to the interference of two resonances, and to the time of flight of
an unstable particle.

To finish this section, we would like to comment on the important issue 
of non-uniqueness of the time operator. Indeed, if $A$ is an invariant 
of the motion, $[H,A]=0$, then the operator $T'\equiv T+\alpha A$ also 
canonically commutes with $H$, where $\alpha$ is a dimensionful constant
that makes $\alpha A$ have dimensions of time. In addition, a given Hamiltonian 
has several energy representations that, although unitarily equivalent, 
are physically nonequivalent~\cite{TRIESTE,HEGERFELDT}. One can use the 
energy representation associated with the ``in'' Lippmann-Schwinger 
eigenfunctions $\langle r|E^+\rangle$~\cite{TRIESTE,HEGERFELDT}, the
one associated with the ``out'' Lippmann-Schwinger eigenfunctions
$\langle r|E^-\rangle$~\cite{TRIESTE,HEGERFELDT}, the one associated with 
the regular solution $\langle r|E\rangle$ of the Schr\"odinger 
equation~\cite{TRIESTE,EXPLAEIGEN}, or the one
associated with the eigenfunctions $\langle r|E_{\Theta} \rangle$ that are 
time-reversal invariant~\cite{HEGERFELDT}. Thus, we may associate many 
time operators with a given Hamiltonian,
and one must select the most appropriate time operator for
the situation at hand. Some selection criteria can be found in
Ref.~\cite{HEGERFELDT}. The reason why we have selected the energy 
representation associated with the ``out'' Lippmann-Schwinger
eigenfunctions is that, as explained in Ref.~\cite{TRIESTE}, such
``out'' energy representation incorporates the final (or detection) boundary 
conditions of a scattering experiment, by contrast to the initial 
(or preparation) conditions of $\langle r|E^+\rangle$.

\section{The Time Representation of a Gamow state}
\setcounter{equation}{0}
\label{sec:timerepGS}

Although the rule~(\ref{dr1})-(\ref{dendindd2}) does not 
rely on the Gamow states, it is nevertheless enlightening to see what such 
states tell us about such rule.

As shown in Ref.~\cite{NPA08}, the Gamow state 
$u(r;z_{\rm R})=\langle r|z_{\rm R}^-\rangle$
associated with a resonant energy $z_{\rm R}=E_{\rm R}-\rmi \Gamma _{\rm R}/2$ has
the following expression in the ``out'' energy representation:
\begin{equation}
     \langle ^-E|z_{\rm R}^- \rangle = 
            \rmi \sqrt{2\pi \, } {\cal N}_{\rm R} \,
      {\delta} (E-z_{\rm R})  \, ,
      \label{-ErepGk}
\end{equation}
where $\delta (E-z_{\rm R})$ is the complex delta function and 
${\cal N}_{\rm R}^2=\rmi \, {\rm res}[S(z)]_{z=z_{\rm R}}$ is Zeldovich's
normalization factor. The time representation of the Gamow state is easily
obtained by Fourier transforming Eq.~(\ref{-ErepGk}):
\begin{eqnarray}
     u(t;z_{\rm R}) &=& 
        \langle t|z_{\rm R}^- \rangle 
       = 
     \int_0^{\infty} \rmd E \, \langle t| E^-\rangle 
            \langle ^-E|z_{\rm R}^- \rangle  \nonumber \\  
      & =&
     \int_0^{\infty} \rmd E \, \frac{1}{\sqrt{2\pi \hbar}}\rme^{-\rmi Et/\hbar}
               \, \rmi \sqrt{2\pi} {\cal N}_{\rm R} \delta (E-z_{\rm R})
               \nonumber \\ 
      &=& \frac{\rmi {\cal N}_{\rm R}}{\sqrt{\hbar}} 
           \rme ^{-\rmi z_{\rm R}t/\hbar}  \, .
      \label{-ErepGktime}
\end{eqnarray}
Hence,
\begin{equation}
     R(t)=|u(t;z_{\rm R})|^2 = \frac{|{\cal N}_{\rm R}|^2}{\hbar} 
           \rme ^{-\Gamma _{\rm R}t/\hbar}  \, 
             \qquad \text{(for a Gamow state)}. 
      \label{-ErepGktimesq}
\end{equation}
Thus, the time representation of the Gamow states yields the 
exponential law and therefore describes the 
exponential region of quantum decay. It should be noted that this decay 
rate is not exactly the same as the decay rate associated with 
the probability of Eq.~(\ref{ndamstand}):
\begin{equation}
       R_{P}(\tau )=-\frac{\rmd P(\tau )}{\rmd \tau} =
             \frac{\Gamma _{\rm R}}{\hbar}\rme ^{-\Gamma _{\rm R}\tau /\hbar} \, .
\end{equation}

\section{A single-resonance system}
\setcounter{equation}{0}
\label{sec:single-resys}

As is well known, quantum mechanics predicts deviations from exponential
decay. Such deviations occur because the Gamow state cannot be prepared
experimentally: All that can be prepared is a
square-integrable wave function $\varphi (t)$. When one resonance is dominant
and we can approximate $\varphi (t)$ by the Gamow state
of the resonance, then one can say
that for all practical purposes the decay is purely exponential and the
Gamow state is the wave function describing quantum decay. It would be 
therefore interesting to see what is the exact expression of the 
decay rate~(\ref{dr1}) when one, and only one, resonance needs to be taken into 
account. In such a case, we can assume that the $S$ matrix has one, and only
one, pole at the resonant energy
$z_{\rm R}=E_{\rm R}-\rmi \Gamma _{\rm R}/2$. Because
$S(E)$ has only one pole and because it is unitary, the residue of $S(E)$ at
$E=z_{\rm R}$ is given by
\begin{equation}
       {\rm res}[S(z)]_{z=z_{\rm R}} = -\rmi \hskip0.2mm \Gamma _{\rm R} \, 
          \qquad \text{(one resonance only)}.
\end{equation}
Hence, $|{\cal N}_{\rm R}|^2=\Gamma _{\rm R}$ and Eq.~(\ref{-ErepGktimesq})
becomes
\begin{equation}
     R(t)=|u(t;z_{\rm R})|^2 =  \frac{\Gamma _{\rm R}}{\hbar} 
           \rme ^{-\Gamma _{\rm R}t/\hbar}= \frac{1}{\tau _{\rm R}} 
           \rme ^{-t/\tau_{\rm R}} \,  \hskip1cm 
           \text{(a single-resonance system)}. 
      \label{-GS1redr}
\end{equation}
If initially there are $N_0$ resonances of the same energy 
$z_{\rm R}=E_{\rm R}-\rmi \Gamma _{\rm R}/2$, the decay rate is
\begin{equation}
        R(t)=N_0|u(t;z_{\rm R})|^2 = \frac{N_0}{\tau _{\rm R}} 
           \rme ^{-t/\tau_{\rm R}} \,  \hskip1cm 
           \text{($N_0$ copies of a single-resonance system)}.  
      \label{-GS1redrN}  
\end{equation}
Comparison of Eqs.~(\ref{-GS1redrN}) and (\ref{decayrate}) shows that,
when only one resonance needs to be taken into account, 
the Gamow state in the time representation yields the exponential
law and the correct initial decay rate:
$R_0=N_0\lambda =\frac{N_0}{\tau _{\rm R}}$.
If we now plug Eq.~(\ref{-GS1redrN}) into Eq.~(\ref{numberofpar}), we 
obtain Eq.~(\ref{numofprtime}). 
Thus, when a quantum system can be approximated by a lone resonance and
its wave function can be approximated by a lone Gamow state, the
time representation of the Gamow state provides a complete quantum-mechanical 
description of the phenomenology of exponential decay. In addition, 
the probability
density associated with the Gamow state is automatically normalized to $1$,
\begin{equation}
        \int_0^{\infty} \rmd t \, |u(t;z_{\rm R})|^2 = 
         \int_0^{\infty} \rmd t \, \frac{1}{\tau _{\rm R}} 
           \rme ^{-t/\tau_{\rm R}} = 1 \, .
\end{equation}


\section{The survival probability vs.~the non-decay
probability}
\setcounter{equation}{0}
\label{sec:survvsdecay}

Quantum decay is usually analyzed by way of the survival
probability, see for example Refs.~\cite{KHALFIN,MUGA1,MUGA2,GASTON} and 
references therein. The survival probability is given by 
$p_{\rm s}(\tau)=|a_{\rm s}(\tau )|^2$, 
where $a_{\rm s}(\tau)$ is the survival amplitude~\cite{EXPLATRS},
\begin{equation}
 a_{\rm s}(\tau) = \langle \varphi|\rme ^{-\rmi H\tau/\hbar}|\varphi \rangle     
     = \int_0^{\infty} \rme ^{-\rmi E\tau/\hbar} |\varphi (E)|^2 \rmd E \, .
   \label{suram}
\end{equation}
By contrast, in the time representation, the non-decay probability amplitude is
\begin{equation}
    {\cal A}(t)\equiv \varphi(t)=\frac{1}{\sqrt{2\pi \hbar}}
               \int_0^{\infty} \rme ^{-\rmi Et/\hbar} \varphi (E) \rmd E \, .
            \label{nda}
\end{equation}
Comparison of Eqs.~(\ref{suram}) and (\ref{nda}) shows several
differences between $a_{\rm s}(\tau)$ and ${\cal A}(t)$.
First, the survival amplitude is dimensionless, whereas the
non-decay amplitude has dimensions of 1/$\sqrt{{\rm time}}$. Second, 
the survival amplitude is the Fourier transform of the absolute value 
squared of the wave function in the energy representation,
whereas the non-decay amplitude is the Fourier transform of
the wave function in the energy representation. Third, in $a_{\rm s}(\tau)$ time 
appears as a parameter, whereas in ${\cal A}(t)$ time appears as a 
random variable. Fourth, as will be further discussed
in Section~\ref{sec:continuousme},
${\cal A}(t)$ seems more suitable to 
situations in which the system is monitored 
continuously, whereas $a_{\rm s}(\tau)$ seems more suitable to situations
where the system evolves freely until the instant $\tau$, at which instant an
instantaneous
measurement is made. Fifth, the decay rate $\dot{p}_{\rm s}(\tau)$ 
associated with $a_{\rm s}(\tau)$ is always zero at 
$\tau =0$~\cite{KHALFIN}, whereas the 
decay rate $R(t)=|\varphi (t)|^2$ is not necessarily zero at $t=0$, as shown in
the Appendix by way of an example. Thus, contrary to $\dot{p}_{\rm s}(\tau)$,
the decay rate $R(t)=|\varphi (t)|^2$ does in general not 
exhibit the Zeno effect.

There are, however, some analogies between $a_{\rm s}(\tau )$ and 
${\cal A}(t)$: Both $a_{\rm s}(\tau )$ and ${\cal A}(t)$ yield the exponential
decay law when the wave function is the Gamow state, and both yield
deviations from exponential decay when the wave function is a properly
normalized wave function $\varphi$. Thus, from a physical point of view, both 
$a_{\rm s}(\tau )$ and ${\cal A}(t)$ can describe quantum decay, and one
should choose one over the other depending on whether the unstable system is
monitored continuously.

\section{Interference of two resonances in the time representation}
\setcounter{equation}{0}
\label{sec:gsianom}

\subsection{The GSI anomaly}

In 2008, Litvinov {\it et al.}~\cite{GSI} observed that K-shell
electron capture (EC) decay rates of Hydrogen-like $^{140}{\rm Pr}^{58+}$
and $^{142}{\rm Pm}^{60+}$ ions,
\begin{eqnarray}
          ^{140}{\rm Pr}^{58+} \to \ ^{140}{\rm Ce}^{58+} + \nu _{\rm e} \, ,
                     \label{reaction1} \\
          ^{142}{\rm Pm}^{60+} \to \ ^{142}{\rm Nd}^{60+} + \nu _{\rm e} \, , 
                      \label{reaction2}
\end{eqnarray}
show an oscillatory modulation superimposed on the exponential decay. The 
decay rate of the GSI anomaly has been fitted with the following 
equation~\cite{GSI}:
\begin{equation}
             \frac{\rmd N_{\rm EC}(t)}{\rmd t} =
                  N_0 \rme ^{-\lambda t}\lambda _{\rm EC}
                   \left(  1 +a \cos (\omega t +\phi)\right) \, ,
            \label{GSI-fit}
\end{equation}
where $N_{\rm EC}$ is the number of daughter ions 
$^{140}{\rm Ce}^{58+}$ and $^{142}{\rm Nd}^{60+}$, $N_0$ is the number of
Hydrogen-like mother ions $^{140}{\rm Pr}^{58+}$
and $^{142}{\rm Pm}^{60+}$, the amplitude $a \simeq 0.20$, and the
period $T=2\pi/\omega \simeq 7$~seconds. The data of~\cite{GSI} were 
obtained by continuously monitoring the decay of individual 
atoms~\cite{CONTI}.

There are 
several theoretical proposals that attempt to explain the oscillations
of the GSI anomaly: Refs.~\cite{LIPKIN,IVANOV,FABER,KLEINERT} 
use neutrino oscillations; Refs.~\cite{GIUNTI1,GIUNTI2,GIUNTI3,KIENERT,MERLE}
use the interference of two mass 
eigenstates; Ref.~\cite{GAL} uses the neutrino spin precession in the
static magnetic field of the storage ring; Ref.~\cite{GIACOSA} uses a 
truncated Breit-Wigner distribution with an energy-dependent width.

Similarly to Refs.~\cite{GIUNTI1,GIUNTI2,GIUNTI3,KIENERT,MERLE}, we 
are going to make the assumption that 
the oscillations of the GSI anomaly are the result of the interference
of two mass eigenstates. Because Litvinov {\it et al.}~\cite{GSI} 
continuously monitored the ions~\cite{CONTI}, we are going to express the decay 
rates of $^{140}{\rm Pr}^{58+}$ and $^{142}{\rm Pm}^{60+}$ in terms of the 
wave function in the time representation as 
in Eqs.~(\ref{dr1}) and (\ref{drn}).

Before proceeding with the time-representation description of the GSI anomaly,
we would like to note that the following results will not explain why 
the GSI anomaly actually occurs, that is, why 
Hydrogen-like $^{140}{\rm Pr}^{58+}$ and $^{142}{\rm Pm}^{60+}$ ions must 
have two resonances that interfere to produce an oscillation superimposed
on exponential decay. 
What the present paper will show is
that if the oscillations of the GSI anomaly were due to the interference of two
resonances, then such interference should be analyzed in the time 
representation. 

When $\varphi (t)$ can be approximated by a Gamow state, 
Eqs.~(\ref{dr1}) and~(\ref{drn}) lead to the exponential law. However, 
since in Eq.~(\ref{GSI-fit}) we have an
oscillation superimposed on the exponential decay, it seems natural
to approximate the wave function $\varphi (t)$ by a superposition
of two Gamow states in the time representation,
\begin{equation}
       \varphi (t) = \langle t|\varphi \rangle 
                   \equiv 
             b_1 \langle t|z_1\rangle + b_2 \langle t|z_2\rangle  
                   = 
             b_1 c_1\rme ^{-\rmi z_1t/\hbar} +b_2 c_2\rme ^{-\rmi z_2t/\hbar}
                      \, ,
           \label{expansion1} 
\end{equation}
where $z_i=E_i-\rmi \Gamma _i/2$, 
$c_i=\frac{\rmi {\cal N}_{i}}{\sqrt{\hbar}}$,
${\cal N}_{i}^2=\rmi \, {\rm res}[S(z)]_{z=z_i}$, $b_i$ are the
mixing coefficients, and $i=1$, $2$. The 
absolute value squared of~(\ref{expansion1}) yields the following
single-particle decay rate:
\begin{eqnarray}
       |\varphi (t)|^2 &=&  
              |b_1c_1\rme ^{-\rmi z_1t/\hbar} + b_2c_2\rme ^{-\rmi z_2t/\hbar}|^2 
                        \nonumber \\
           &=& |b_1|^2 |c_1|^2\rme ^{-\Gamma _1t/\hbar} + 
               |b_2|^2 |c_2|^2\rme ^{-\Gamma _2t/\hbar} +
                2|b_1||c_1| |b_2| |c_2| \rme ^{- \frac{(\Gamma _1 +\Gamma _2)t}{2\hbar}}  
                 \cos \left( \frac{\Delta E \, t }{\hbar} +\delta \right) ,
                   \nonumber \\               
            & \quad & \quad
                \label{timeintef2re}
\end{eqnarray}
where $\Delta E=E_1-E_2$,
$c_i =|c_i| \rme ^{-\rmi \delta _i}$, $b_i =|b_i| \rme ^{-\rmi \delta '_i}$,
and $\delta=\delta _1-\delta _2+\delta '_1 -\delta ' _2$. 
Equation~(\ref{timeintef2re}) is the most general form for the interference
of two Gamow states in the time representation. Such interference will
in general produce an exponential decay coupled to an oscillation between
the two modes of decay. If the lifetimes of the resonances are 
not the same, then the oscillation will be damped until
the first resonance has decayed, and after that the decay will 
be essentially exponential through the longer-lived resonance.

Since the amplitude of the oscillation of the GSI anomaly is not damped,
in order to reproduce the decay rate~(\ref{GSI-fit}), we are going to assume
that the decay widths of the two resonances $z_1$ and $z_2$ are the same
(or, equivalently, that their lifetimes are the same):
\begin{equation}
     \Gamma _1 =\Gamma _2 =\Gamma \, . 
        \label{eqgamma}
\end{equation}
Substituting Eq.~(\ref{eqgamma}) into the single-particle
decay rate~(\ref{timeintef2re}) yields
\begin{equation}
       |\varphi (t)|^2 = \rme ^{-\Gamma t/\hbar} (|b_1|^2|c_1|^2 + |b_2|^2|c_2|^2)
                \left[ 1+ \frac{2|b_1||c_1||b_2||c_2|}{|b_1|^2|c_1|^2 + 
                 |b_2|^2|c_2|^2}
                 \cos \left( \frac{\Delta E \, t }{\hbar} +\delta \right) 
                      \right]  .           \label{timeintef2re2}
\end{equation}
By combining Eqs.~(\ref{decminusin}), (\ref{drn}) and~(\ref{timeintef2re2}),
we obtain the rate at which $N_0$ unstable nuclei decay 
through two resonances $z_1$ and $z_2$ of the same width,
\begin{equation}
       \frac{\rmd N_{\rm EC}(t)}{\rmd t}=
           N_0\rme ^{-\Gamma t/\hbar} (|b_1|^2|c_1|^2 + |b_2|^2|c_2|^2)
          \left[ 1+ \frac{2|b_1||c_1| |b_2||c_2|}{|b_1|^2|c_1|^2 + |b_2|^2|c_2|^2}
                 \cos \left( \frac{\Delta E \, t }{\hbar} +\delta \right) 
                      \right]   .  
               \label{timeintef2re3}
\end{equation}
Comparison of Eqs.~(\ref{timeintef2re3}) and (\ref{GSI-fit}) shows that
those two equations are identical if we make the following identifications:
$\lambda =\Gamma /\hbar$, $\lambda _{\rm EC}= |b_1|^2|c_1|^2 + |b_2|^2|c_2|^2$, 
$\omega = \frac{\Delta E}{\hbar}$, $\phi = \delta$, and
$a = \frac{2|b_1||c_1| |b_2||c_2|}{|b_1|^2|c_1|^2 + |b_2|^2|c_2|^2}$. Thus, 
the GSI anomaly can be interpreted as the interference 
of two resonances in the time representation.

\subsection{Description of the GSI anomaly in terms of the survival probability}

In this
section, we are going to see that the decay rate of the survival probability
of two interfering resonances has mathematical similarities to and physical 
differences from the decay rate of Eq.~(\ref{timeintef2re3}).

Let us consider a wave function that can be approximated
by a coherent superposition of two Gamow states with amplitudes
$b_1$ and $b_2$,
\begin{equation}
       |\varphi \rangle \equiv b_1 |z_1\rangle + b_2 |z_2\rangle \, .
          \label{surviaexp}
\end{equation}
The survival amplitude of such state is~\cite{FN1}
\begin{equation}
       a_{\rm s}(\tau ) =  
                \langle \varphi |\rme ^{-\rmi H\tau /\hbar}| \varphi \rangle  
              = |b_1|^2\rme ^{-\rmi z_1\tau /\hbar} + 
                |b_2|^2\rme ^{-\rmi z_2\tau /\hbar} \, , 
\end{equation}
where we have assumed that the Gamow states are normalized such that
$\langle z_i|z_j\rangle =\delta _{ij}$. The survival probability is given by
\begin{equation}
           p_{\rm s}(\tau ) = |a_{\rm s}(\tau )|^2   
              = |b_1|^4\rme ^{-\Gamma _1\tau/\hbar} + 
                |b_2|^4\rme ^{-\Gamma _2\tau/\hbar} +
                2|b_1|^2 |b_2|^2 \rme ^{- \frac{(\Gamma _1 +\Gamma _2)\tau}{2\hbar}}  
                 \cos \left( \frac{\Delta E \, \tau }{\hbar} \right) .
           \label{survpro}
\end{equation}
Since Litvinov {\it et al.}~\cite{GSI} did not measure a probability but 
rather a decay rate, we need to calculate the decay rate associated
with $p_{\rm s}(\tau )$ when initially there are $N_0$ unstable ions,
\begin{eqnarray}
       \frac{\rmd N_{\rm s}(\tau )}{\rmd \tau} &=&
          -N_0\frac{\rmd p_{\rm s}(\tau )}{\rmd \tau}  \nonumber \\
           &=&
           N_0 \frac{\Gamma }{\hbar} \rme ^{-\Gamma \tau/\hbar} 
           (|b_1|^4 + |b_2|^4)
         \left(1+\frac{2|b_1|^2 |b_2|^2}{|b_1|^4 + |b_2|^4}\left[        
      \cos \left( \frac{\Delta E \, \tau }{\hbar} \right)
   + \frac{\Delta E}{\Gamma}  
                 \sin \left( \frac{\Delta E \, \tau }{\hbar}\right)
            \right] \right) ,  \nonumber \\
                \label{timeintef2re-suvr-eq}
\end{eqnarray}
where we have assumed that $\Gamma_1=\Gamma_2=\Gamma$. If we 
define $A\rme ^{-\rmi \psi}\equiv 1  +\rmi \frac{\Delta E }{\Gamma}$, 
then Eq.~(\ref{timeintef2re-suvr-eq}) can be written as
\begin{equation}
       \frac{\rmd N_{\rm s}(\tau )}{\rmd \tau} =N_0
     \frac{\Gamma }{\hbar} \rme ^{-\Gamma \tau/\hbar} 
           (|b_1|^4 + |b_2|^4)
         \left(1+\frac{2|b_1|^2 |b_2|^2}{|b_1|^4 + |b_2|^4}A 
       \cos \left( \frac{\Delta E \, \tau }{\hbar} +\psi \right) \right) .
                  \label{timeintef2re-suvr-simpl}
\end{equation}
Comparison of Eqs.~(\ref{timeintef2re-suvr-simpl}) and~(\ref{GSI-fit})
shows that the survival probability can also account for the GSI
anomaly if we define $\lambda _{\rm EC}= \frac{\Gamma}{\hbar}(|b_1|^4 + |b_2|^4)$,
$\lambda =\frac{\Gamma}{\hbar}$, $a=\frac{2|b_1|^2 |b_2|^2}{|b_1|^4 + |b_2|^4}A$,
and $\phi =\psi$. 

Mathematically, Eqs.~(\ref{timeintef2re3}) and (\ref{timeintef2re-suvr-simpl})
both consist of an oscillation superimposed on exponential decay. Physically, 
however, Eqs.~(\ref{timeintef2re3}) and 
(\ref{timeintef2re-suvr-simpl}) have several differences. First, 
the coefficients needed by Eqs.~(\ref{timeintef2re3}) and 
(\ref{timeintef2re-suvr-simpl}) in order to account for 
Eq.~(\ref{GSI-fit}) are different and, in principle, experimentally 
distinguishable. For example, if we were able to vary the mixing coefficients 
$b_1$ and $b_2$ at will, Eqs.~(\ref{timeintef2re3}) and 
(\ref{timeintef2re-suvr-simpl}) would yield distinguishable 
decay rates. Second, in Eq.~(\ref{timeintef2re3}) time appears 
as a random variable, whereas in Eq.~(\ref{timeintef2re-suvr-simpl}) 
time appears as a parameter. Third, as we will further
discuss in Section~\ref{sec:continuousme}, Eq.~(\ref{timeintef2re3}) is 
based on the assumption that the system is monitored continuously 
(as is the case of the GSI anomaly), 
whereas Eq.~(\ref{timeintef2re-suvr-simpl}) assumes that the system evolves 
freely up to the instant $\tau$, at which instant an instantaneous measurement 
is made.

\subsection{The Quantum-Beat description of the GSI anomaly}

Assuming that the GSI anomaly is due to the interference of two 
resonances in the time representation is very close to assuming that such
anomaly is due to the ``quantum beats'' of two exponentially decaying 
mass eigenstates~\cite{GIUNTI1,GIUNTI2,GIUNTI3,KIENERT,MERLE}. It seems 
therefore
pertinent to compare the quantum-beat approach with Eq.~(\ref{timeintef2re3}).

Instead of the survival probability $p_{\rm s}(\tau)$, in the quantum-beat 
approach one obtains the following transition probability of electron 
capture at time $\tau$~\cite{GIUNTI1}:
\begin{equation}
    P_{\rm QB}(\tau)=\bar{P} \, \rme ^{-\Gamma \tau/\hbar} 
                \left[ 1+ b
                 \cos \left( \frac{\Delta E \, \tau }{\hbar} +\delta \right) 
                      \right]   ,           \label{timeintef2re2Giun}
\end{equation}
where $\bar{P}$, $b$ and $\delta$ are constants. The decay rate 
associated with $P_{\rm QB}(\tau)$ is given by
\begin{equation}
       \frac{\rmd N_{\rm QB}(\tau )}{\rmd \tau} =
             N_0\bar{P} \rme ^{-\Gamma \tau/\hbar}
           \left[ \frac{\Gamma}{\hbar}
                \left( 1+ b
                 \cos \left( \frac{\Delta E \, \tau }{\hbar} +\delta \right) 
                      \right)  
              + b \frac{\Delta E}{\hbar} 
               \sin \left( \frac{\Delta E \, \tau }{\hbar} +\delta \right)
              \right]  .
                \label{timeintef2re-Giunti-dec}
\end{equation}
If we define $B\rme ^{-\rmi \psi}=b+\rmi b\frac{\Delta E}{\Gamma}$, 
Eq.~(\ref{timeintef2re-Giunti-dec}) becomes
\begin{equation}
       \frac{\rmd N_{\rm QB}(\tau )}{\rmd \tau} =
             N_0\bar{P}\frac{\Gamma}{\hbar} \rme ^{-\Gamma \tau/\hbar}
           \left[ 
                1+ B
            \cos \left( \frac{\Delta E \, \tau }{\hbar} +\delta +\psi \right) 
                         \right]  .
                \label{timeintef2re-Giunti-dec-l}
\end{equation}
Thus, the quantum-beat approach accounts for Eq.~(\ref{GSI-fit}) but
with different (and, in principle, experimentally distinguishable)
coefficients than those of Eq.~(\ref{timeintef2re3}). In addition, similarly to
the survival probability, the quantum-beat approach treats time as
a parameter and implicitly assumes that the system is not monitored
continuously.

It should be noted that molecular and atomic fluorescence quantum beats 
are studied using 
an equation similar to Eq.~(\ref{timeintef2re2Giun}), see for example
Ref.~\cite{HUBER} . Therefore, when the system is monitored continuously 
in an atomic or molecular quantum-beat experiment (as, for example, 
in Ref.~\cite{HALL}), the theoretical description of
quantum beats may have to be done in the time representation~\cite{QBHALL}.

\section{Neutral-meson decay}
\setcounter{equation}{0}
\label{sec:kaon}

Similarly to Hydrogen-like $^{140}{\rm Pr}^{58+}$ and $^{142}{\rm Pm}^{60+}$ ions,
the decays of $K$ and $B$ mesons exhibit oscillations superimposed on 
exponential decay. In this section, we are going to see how the oscillations of 
neutral mesons can be described in the time representation 
(for a somewhat related approach, see Refs.~\cite{COURBAGE,DURT}).

In the Lee-Oehme-Yang model of the kaon system, 
the mass operator has two mass eigenstates $\klong$ and $\kshort$ with complex 
eigenvalues $z_{\rm L}= m_{\rm L}c^2-\rmi \Gamma _{\rm L}/(2\hbar)$ and 
$z_{\rm S}= m_{\rm S}c^2-\rmi \Gamma _{\rm S}/(2\hbar)$. Thus, $\klong$ and 
$\kshort$ are two Gamow states. The $\knot$ and $\knotbar$ can be written 
in terms of such Gamow states as follows
\begin{eqnarray}
        &&\knot = \frac{1}{{\sqrt{2}}} \left( \kshort + \klong \right) \, ,
                 \label{knot} \\
        &&\knotbar = \frac{1}{{\sqrt{2}}} \left( \kshort - \klong \right) \, .
                 \label{knotbar}
\end{eqnarray}
The time representation of Eqs.~(\ref{knot}) and~(\ref{knotbar}) reads as
\begin{eqnarray}
        &&\varphi_{\knot} (t) = \frac{1}{{\sqrt{2}}} 
                          \left(c_{\rm S}\rme ^{-\rmi z_{\rm S}t/\hbar} + 
                        c_{\rm L}\rme ^{-\rmi z_{\rm L}t/\hbar} \right) \, , 
                 \label{knot-t} \\
        &&\varphi _{\knotbar}(t) = \frac{1}{{\sqrt{2}}} 
                          \left(c_{\rm S}\rme ^{-\rmi z_{\rm S}t/\hbar} - 
                        c_{\rm L}\rme ^{-\rmi z_{\rm L}t/\hbar} \right) \, , 
                 \label{knotbar-t}
\end{eqnarray}
where $c_i=\frac{\rmi {\cal N}_{i}}{\sqrt{\hbar}}$,
${\cal N}_{i}^2=\rmi \, {\rm res}[S(z)]_{z=z_i}$, and $i={\rm S}$, ${\rm L}$. We 
can derive the decay rates for the neutral kaons in complete analogy to 
the way we derived the decay rate~(\ref{timeintef2re3}) from 
Eq.~(\ref{expansion1}). If we start off with a pure $\knot$ beam 
at $t=0$, then the decay rate is given by
\begin{equation}
       \frac{\rmd N_{\knot}(t)}{\rmd t} =
                  \frac{N_0}{2}\left[
                    |c_{\rm S}|^2\rme ^{-\Gamma _{\rm S}t/\hbar} + 
                |c_{\rm L}|^2\rme ^{-\Gamma _{\rm L}t/\hbar} +
                2|c_{\rm S}| |c_{\rm L}| 
                   \rme ^{- \frac{(\Gamma _{\rm S} +\Gamma _{\rm L})t}{2\hbar}}  
                 \cos \left( \frac{\Delta E \, t }{\hbar} +\delta \right)
                  \right]  .
                \label{knotdr}
\end{equation}
If we start off with a pure $\knotbar$ beam at $t=0$, then the decay rate
is
\begin{equation}
       \frac{\rmd N_{\knotbar}(t)}{\rmd t} =
                  \frac{N_0}{2}\left[
                    |c_{\rm S}|^2\rme ^{-\Gamma _{\rm S}t/\hbar} + 
                |c_{\rm L}|^2\rme ^{-\Gamma _{\rm L}t/\hbar} -
                2|c_{\rm S}| |c_{\rm L}| 
                   \rme ^{- \frac{(\Gamma _{\rm S} +\Gamma _{\rm L})t}{2\hbar}}  
                 \cos \left( \frac{\Delta E \, t }{\hbar} +\delta \right)
                  \right]  .
                \label{knotdrbar}
\end{equation}
Equations~(\ref{knotdr}) and~(\ref{knotdrbar}) have the same form as those
used in the literature to calculate the time dependence of the decay rates of
$\knot$ and $\knotbar$ (see, for example,
Eqs.~(7.56), (7.57) and (7.64) in Ref.~\cite{PERKINS}, or Eq.~(3) in 
Ref.~\cite{CPLEAR}). A similar procedure can be applied to the decay 
of B mesons~\cite{ABE}. 

Thus, if in an experiment measuring the decay of neutral mesons 
the particles were monitored continuously, one could describe 
such decay in a time representation that is obtained by Fourier transforming
the representation where the mass operator is diagonal~\cite{DISC}.


\section{Continuous measurements and the Zeno effect}
\setcounter{equation}{0}
\label{sec:continuousme}

In this section, we are going compare the
procedure to measure the survival probability $p_{\rm s}(\tau )$ with
the procedure to measure the non-decay probability ${\cal P}(t)$. From
such comparison we will conclude that ${\cal P}(t)$ is more
suitable than $p_{\rm s}(\tau)$ to model the decay of a particle
that is monitored continuously.

\subsection{Measurement of the survival probability}

Let a quantum system be initially prepared in the state $\varphi$. It
is usually assumed that the probability that the system remains in the 
state $\varphi$ 
after a time $\tau _0$ is given by the survival probability, 
$p_{\rm s}(\tau _0) = 
|\langle \varphi |\rme^{-\rmi H\tau_0 /\hbar}|\varphi \rangle |^2$. If
we want to measure the survival probability at $\tau =\tau _0$, we 
need to prepare the system in the state $\varphi$ at $\tau =0$,
let the system evolve unmonitored until $\tau =\tau _0$, and finally make
an instantaneous measurement at $\tau =\tau _0$. We then say that
we have measured the observable $P=|\varphi \rangle \langle \varphi |$
on the state $\rme ^{-\rmi H\tau _0/\hbar}\varphi$.

If we want to measure the survival probability at time $\tau =2\tau _0$, 
we prepare the system in the state $\varphi$ at $\tau =0$, let 
the system evolve unmonitored till $\tau =2\tau _0$, and finally, without 
performing any measurement prior to $\tau =2\tau _0$, 
perform an instantaneous measurement at $\tau =2\tau _0$. We then say that
we have measured the observable $P=|\varphi \rangle \langle \varphi |$
on the state $\rme ^{-\rmi H2\tau _0/\hbar}\varphi$.

Thus, in order to measure $p_{\rm s}(\tau )$, we must perform a different
experiment for each instant of time $\tau$. All these measurements are 
projective measurements, since the observable we are measuring is 
the projection $P= |\varphi \rangle \langle \varphi |$.

\subsection{Pulsed and continuous measurements}

Let us now assume that we perform a pulsed measurement, that is, we
prepare the system in the sate $\varphi$ at $\tau =0$, and 
then measure the probability
that the system remains in the state $\varphi$ 
at times $\tau _0$, $2\tau _0$, $3\tau _0$,
and so on. Due to the reduction postulate, the probabilities
$p_{\rm pulsed}(n\tau _0)$ that we will obtain at times $n\tau_0$, 
$n=2, 3, \ldots$, will in general be different from the survival probabilities
$p_{\rm s}(n\tau_0)$, $n=2,3,\ldots$. It has been found both
theoretically~\cite{SIEGERT,SUDARSHAN,DEGASPERIS,FLEMING,
FONDAREV,KRAUS,SUDBERY,JOOS,KOFMAN,SCHULMAN,PASCA1,PASCA2,EXPLAREF} 
and experimentally~\cite{ITANO,RAIZEN,STREED} that the 
probabilities $p_{\rm pulsed}(n\tau _0)$
can be larger or smaller than $p_{\rm s}(n\tau_0)$ . When they are larger,
we say that the evolution (or decay, in the case of an unstable system) 
is hindered by the measurement, and we refer to it as the Zeno 
effect. When they are smaller, we say that the evolution of the system is 
sped up, and we refer to it as the anti-Zeno effect. The Zeno effect 
was observed experimentally for the first time
in Ref.~\cite{ITANO} for Rabi oscillations, and both the Zeno and anti-Zeno 
effects were first observed for a decaying system in Ref.~\cite{RAIZEN}.



Finally, let us consider the case in which the pulsed measurement is so
frequent that it can be assumed to be a continuous measurement. In such a
case, due to the reduction postulate, the measurements continuously collapse
the wave function to the state $\varphi$, and the evolution slows down to 
a stop. This case, which was first observed for Rabi oscillations
in Ref.~\cite{ITANO}, is usually referred to either as the Zeno effect, as
the ``watchdog effect,'' or as the ``watched pot never boils'' effect. 


\subsection{Measuring the non-decay probability vs.~measuring the
survival probability}

We have seen that the measurement of the survival probability is a 
projective measurement in which
the observable $|\varphi \rangle \langle \varphi |$ is measured on the 
state $\rme ^{-\rmi H\tau /\hbar}\varphi$. A position-representation analog is 
the measurement of $|x\rangle \langle x|$ on the 
state $\rme ^{-\rmi H\tau /\hbar}\varphi$ to
obtain the probability density $|\varphi (x;\tau)|^2$ that the state 
$\varphi$ is found at position $x$ at time $\tau$. When we measure 
$|\varphi (x;\tau)|^2$ or $p_{\rm s}(\tau )$, we have complete control
over the instant $\tau$ at which the measurement is performed. The detectors
are turned off prior to the instant $\tau$, and when such instant arrives, 
we perform an instantaneous measurement. When we measure $p_{\rm s}(\tau)$, 
the question to be answered is, 
``what is the probability that the system has not decayed at time $\tau$?''
When we measure $|\varphi (x;\tau)|^2$, the question to be answered
is, ``what is the probability density that the system is found at position
$x$ at time $\tau$?''

By contrast, when we measure the probability density (decay rate, in the 
case of a resonance) $|\varphi (t)|^2$, we measure $|t\rangle \langle t|$ 
on the state $\varphi$. Because of Eq.~(\ref{nonorth}), $|t\rangle \langle t|$
is not a projection and therefore the measurement of $|\varphi (t)|^2$
is not a projective measurement but rather a POVM.  

When we measure ${\cal P}(t)$, we have no control over the time at which 
the decay will occur, because the decay occurs at a random time 
(hence the need to promote time to a random variable). When we measure 
${\cal P}(t)$, we need to monitor the decay of the unstable system 
continuously, or else we may miss the moment when it decays. When
we measure ${\cal P}(t)$, the question to be answered is, ``at what
time does the decay event happen, and with what probability?''

Because the POVM $|t\rangle \langle t|$ is associated with 
the continuous random variable $t$, because $t$ is an eigenvalue 
of a time operator, and because the POVM 
$|t\rangle \langle t|$ is generated by a time operator, it 
seems reasonable to 
interpret the measurement of $|t\rangle \langle t|$ on 
$\varphi$ (i.e., the measurement of ${\cal P}(t)$) as 
a continuous measurement.

Actually, continuous measurements are inherent to the nature of many time 
operators, the prototypical example being the time-of-arrival 
operator. When one measures the probability that a particle arrives at 
a given position, one needs to monitor the arrival of the particle at 
all times, or else one may miss the moment when the particle 
arrives~\cite{EXPLATOA}.

\subsection{The GSI experiment}

In Ref.~\cite{GSI}, Litvinov {\it et al.}~continuously measured the
decay of the ions. If Litvinov {\it et al.}~were measuring the survival
probability, by the ``watched pot never boils'' effect, the ions would not 
decay. However, the ions of the GSI experiment eventually 
do decay, and therefore the survival probability does not seem to be the
quantity measured in Ref.~\cite{GSI}. By contrast, the non-decay 
probability ${\cal P}(t)$ seems a natural quantity to analyze the GSI 
anomaly, because ${\cal P}(t)$ can both model continuous measurements and 
account for oscillations superimposed on exponential decay.

There is a definitive test that would allow us to determine whether or not
Litvinov {\it et al.}~measured $p_{\rm s}(\tau)$. If they were able to
measure the decay rate at $t=0$~\cite{EXPLACOOLING}, and if such measurement 
yielded a non-zero initial decay rate, then we would know for sure 
that Litvinov {\it et al.}~cannot possibly be measuring the survival 
probability.

\subsection{Effect of the measurement on the state}

When we perform a projective measurement, the effect of the measurement on
the state is taken into account by the reduction postulate. In addition,
one can model the measuring apparatus by way of a 
quantum-mechanical Hamiltonian, 
see for example Refs.~\cite{KRAUS,SUDBERY,JOOS,SCHULMAN}. 

It would be interesting to also account for the effect of the
measurement of ${\cal P}(t)$ on the state. However,
the measurement of ${\cal P}(t)$ is a POVM, for which there does not seem to 
exist a simple, succinct answer as to what the state is after the measurement, 
or as to how to model the detector with a quantum-mechanical Hamiltonian
(see Ref.~\cite{BUSCH}, and Chapter~3 of Ref.~\cite{PRESKILL}). Thus, the 
results of the present paper simply provide a rule to calculate
the probability for a resonance to decay when such resonance is monitored 
continuously, without explicitly taking the effect of the apparatus into
account.



\section{Time of flight}
\setcounter{equation}{0}
\label{sec:timeofflig}

Times of flight are routinely measured in the lab, and they seem to
have a dynamical character. In this section,
we are going to use the time representation to construct a quantum-mechanical 
description of the time of flight.

Let us assume that an experimenter can measure the time of flight
of an unstable particle by measuring, for example, the length of the trails left
by the particle in a bubble or spark chamber. If 
$v$ is the speed of the particle and $d_i$ is the length of the
trail, then the time of flight is just $t_i=d_i/v$, where
$i$ labels the trails left by the particle in different, successive
experiments. The times
$t_i$ are random, dynamical times, not parametric times over which the
experimenter has complete control. The 
randomness of $t_i$ arises from the seemingly irreducible randomness of 
quantum decay: You cannot predict when an individual particle is going to 
decay, all you can predict is the probability for such a decay to occur.

Let us assume that each time of flight $t_i$ is obtained $N_i$ times when
we repeat the same experiment $N_0$ times (or, equivalently, when we
perform a single experiment with $N_0$ particles that decay independently 
of each other). Because the times $t_i$ are random, the average 
time of flight should 
be given by the mean of the corresponding probability distribution,
\begin{equation}
       \tf = \langle t\rangle =\sum \frac{N_i}{N_0}t_i \equiv \sum p_i t_i \, ,
           \label{timeoff}
\end{equation}
where $p_i$ is the probability to measure $t_i$. In the limit that $N_0$
is very large, we obtain a continuous probability distribution
$p(t)$, $p_i$ tends to $\frac{\rmd p(t)}{\rm d t} \rmd t$, 
and Eq.~(\ref{timeoff}) tends to 
\begin{equation}
       \tf  =\int t \frac{\rmd p(t)}{\rmd t} \rmd t \, .
\end{equation}
By assuming that $p(t)$ coincides with ${\cal P}_{\rm d}(t)$, and by using 
rule~(\ref{dendindd}), we obtain the following expression:
\begin{equation}
       \tf =\int t |{\varphi}(t)|^2 \,  \rmd t \, ;
              \label{timeofflig}
\end{equation}
that is, the time of flight of a particle is just the mean (or first moment) 
of the probability distribution associated with the time representation
of the wave function. In the case of a single-resonance system, 
the wave function is given by the Gamow state. By combining
Eq.~(\ref{-GS1redr}), Eq.~(\ref{timeofflig}), and the 
fact that $\int \rmd x \, x\rme ^{-x} = -\rme ^{-x}(x+1)$, we obtain
\begin{equation}
       \tf =\int_0^{\infty} t |u(t;z_{\rm R})|^2 \, \rmd t = 
           \frac{\hbar}{\Gamma _{\rm R}} = \tau _{\rm R} \, ;
              \label{timeoffligre}
\end{equation}
that is, rule~(\ref{dr1})-(\ref{dendindd2}) implies that the time of flight of 
a resonance is the same 
as its lifetime. Because we have obtained Eqs.~(\ref{timeofflig}) and
(\ref{timeoffligre}) by way of the time representation, we can
say that we have endowed the time of flight with a dynamical 
character~\cite{EXPLATOF}.

\section{Conclusions}
\setcounter{equation}{0}
\label{sec:conclusions}

We have presented an analysis of quantum decay in which the decay
of a single unstable particle is described by a single wave function
in the time representation, $\varphi (t)$. Mathematically, $\varphi (t)$
is the Fourier transform of the wave function in the energy 
representation. The decay rate is given by the absolute value squared of
$\varphi (t)$. Mathematically, the decay rate is just the probability 
distribution generated by the POVM of a time operator. The resulting 
non-decay probability appears as a 
natural replacement for the survival probability in situations where the system
is monitored continuously. When 
the analysis is applied to the Gamow state, one
recovers all the phenomenological features of exponential decay, including
the dynamical role played by time. In addition, $\varphi (t)$ provides
a simple way to model quantum measurements that monitor the
system continuously.

We have applied the analysis to the interference of two resonances,
a phenomenon that occurs in a wide variety
of energy ranges, from atomic and molecular fluorescence quantum beats, 
to neutral mesons, to (possibly) the GSI anomaly. When an unstable system 
can decay through two different resonances, it oscillates between them, and the 
ensuing decay rate is given by an oscillation superimposed on 
the exponential decay. We have argued that when the interfering 
resonances are monitored continuously, as is the case of the GSI 
anomaly,
the decay rate should be given in terms of the wave function in the time
representation, Eq.~(\ref{timeintef2re3}), rather than in terms of
the decay rate of the survival 
probability, Eq.~(\ref{timeintef2re-suvr-simpl}), or in terms of 
the decay rate of Eq.~(\ref{timeintef2re-Giunti-dec-l}). Theoretically, 
Eq.~(\ref{timeintef2re3}) has two main advantages. First, 
it endows time with a dynamical character. Second, it explains 
why the system decays even though it is monitored continuously.

Although mathematically Eqs.~(\ref{timeintef2re3}), 
(\ref{timeintef2re-suvr-simpl}), 
and~(\ref{timeintef2re-Giunti-dec-l}) are very similar, they can 
in principle be distinguished experimentally. In particular, if 
the experiment of Ref.~\cite{GSI} measured the decay rate around $t=0$, 
we could find out whether such experiment should be described by 
Eq.~(\ref{timeintef2re3}) or by Eq.~(\ref{timeintef2re-suvr-simpl}).

We have also introduced an expression for the 
time of flight of a quantum particle as the mean of the probability 
distribution of the wave function in the time representation. From 
such expression there follows that the time of flight
of a resonance coincides with its lifetime.


\section*{Acknowledgment}

The author wishes to thank the participants of the ECT* workshop
``Many-body Open Quantum Systems: From Atomic Nuclei to Quantum Optics,''
Trento, September 2012,
for enlightening discussions, and Gregory Hall 
for enlightening correspondence on Ref.~\cite{HALL}. Special thanks are
due to the referees, whose numerous 
suggestions considerably improved the paper. This 
research has been partially 
supported by Ministerio de Ciencia e Innovaci\'on of Spain under 
project TEC2011-24492.

\appendix
\def\thesection{\Alph{section}}
\section{Appendix}
\setcounter{equation}{0}
\label{sec:appendix-cdf}

Let $\varphi (E) = \sqrt{\frac{2 \alpha}{\hbar}} \rme ^{-E\alpha /\hbar }$, where
$\alpha >0$. The time representation
of $\varphi (E)$ is~\cite{EXPLAMSI}
\begin{equation}
       \varphi (t) = \int_0^{\infty} \rmd E \, \varphi (E) 
        \frac{1}{\sqrt{2\pi \hbar}}\rme ^{-\rmi Et/\hbar} =
         \sqrt{\frac{\alpha}{\pi}} \frac{1}{\alpha +\rmi t} \, . 
            \label{timerewf00}
\end{equation}
The decay rate associated with the time representation is then given by
\begin{equation}
        R(t)=|\varphi (t)|^2 =   
           \frac{\alpha }{\pi(\alpha ^2 +t^2)} \, .
            \label{timerewf099}
\end{equation}

For the same state 
$\varphi (E) = \sqrt{\frac{2 \alpha}{\hbar}} \rme ^{-E\alpha /\hbar }$, the
survival amplitude is
\begin{equation}
 a_{\rm s}(\tau)= \int_0^{\infty} \rme ^{-\rmi E\tau/\hbar} |\varphi (E)|^2 \rmd E
     = \frac{2\alpha}{2\alpha + \rmi \tau} \, .
   \label{suram00}
\end{equation}
The survival probability and its corresponding decay rate are
\begin{equation}
   p_{\rm s}(\tau )=  |a_{\rm s}(\tau)|^2= \frac{4\alpha ^2}{4\alpha ^2 + \tau ^2} 
            \, ,
   \label{suram099}
\end{equation}
\begin{equation}
   \dot{p}_{\rm s}(\tau )= -\frac{8\alpha ^2 \tau }{(4\alpha ^2 + \tau ^2)^2} 
            \, .
   \label{suram097}
\end{equation}
From the comparison of Eq.~(\ref{timerewf099}) with Eq.~(\ref{suram097}) it 
follows that $|\varphi (t)|^2\neq - \dot{p}_{\rm s}(t)$. In particular,
$R(0)=1/(\pi \alpha ) \neq 0$, whereas $\dot{p}_{\rm s}(0)=0$. 

Thus, the wave function
$\varphi (E) = \sqrt{\frac{2 \alpha}{\hbar}} \rme ^{-E\alpha /\hbar }$ 
shows that in general the decay rate of the non-decay probability 
${\cal P}(t)$ is different from the decay rate of the 
survival probability $p_{\rm s}(\tau)$.



\begin{thebibliography}{99}
\bibitem{FN1} Whenever time has a dynamical character, we will denote it
by $t$, but when it plays the role of a parameter, we will denote it by $\tau$.
\bibitem{EXPLAMSI} Strictly speaking, we should attach a ``minus'' sign 
superscript to the wave function, $\varphi \equiv \varphi ^-$, 
in order to specify that we are using the energy representation 
associated with the ``out'' Lippmann-Schwinger equation. In addition,
we should attach a subscript to the wave function, in order to 
indicate whether we are dealing with the position 
(e.g., $\varphi _{\mbox{\tiny pos}}(r)$), the energy 
(e.g., $\varphi _{\mbox{\tiny ene}}(E)$), or the time 
(e.g., $\varphi _{\mbox{\tiny time}}(t)$) representation of $\varphi$, since 
in general the expression of a given wave function $\varphi$ is different in
each representation. However, for the sake of simplicity, we will drop 
such minus superscript, and we will simply write $\varphi (r)$,
$\varphi (E)$ and $\varphi (t)$ to denote the wave function in the 
position, energy and time representations.
\bibitem{EXPLATI} We can nevertheless calculate the parametric time
evolution of $\varphi (t)$, the result of which is
$\rme ^{-\rmi H\tau/\hbar}\varphi (t)=\varphi (t+\tau)$.
\bibitem{CINVES} Because the Hilbert-space spectrum of a
Hermitian Hamiltonian is real, 
it includes only the bound and the scattering spectra, thereby discarding the 
resonant spectrum as unphysical. However, radioactive nuclei and unstable 
elementary particles are physical, and they are included in the
periodic table of the elements and in the particle data table. Therefore,
when a Hermitian Hamiltonian produces resonances, its physical spectrum (which
includes the complex resonant energies) is not the same as its
Hilbert-space spectrum (which does not include the resonant energies).
\bibitem{EXPLAEIGEN} In Ref.~\cite{TRIESTE}, the eigenfunction 
$\langle r|E\rangle$ was called
the standing-wave eigenfunction, which is a misnomer.
\bibitem{EXPLATRS} The wave function $\varphi (E)$ that appears in
Eq.~(\ref{suram}) is usually taken to be the one associated with the
energy representation of the eigenfunctions 
$\langle r|E\rangle$ (see Section~\ref{sec:timerpe}). For simplicity
in the discussion, we are assuming that Eq.~(\ref{suram}) also 
uses the ``out'' energy representation, since similar conclusions would hold
if we used the energy representation of the eigenfunctions 
$\langle r|E\rangle$.
\bibitem{CONTI} Although no real measurement truly monitors the
system continuously, we can consider a measurement to be continuous 
whenever it monitors the system frequently enough. Such seems to be
the case of Ref.~\cite{GSI}, because the ions revolve in the ring
with a period of the order of $10^{-6}$~s and are monitored nondestructively
once per revolution.
\bibitem{QBHALL} Strictly speaking, the quantum beats of Ref.~\cite{HALL}
do not involve decay and therefore do not use Eq.~(\ref{timeintef2re2Giun}).
Nevertheless, since the system is monitored continuously, they may have to
be described in an appropriate time representation.
\bibitem{DISC} Such description is of course constrained to the
Lee-Oehme-Yang effective theory of the kaon system, and it is
not valid within the quantum-field-theory description of such system.
\bibitem{EXPLATOA} Incidentally, we would like to note that 
the ``watched pot never boils'' effect
does not occur in measurements of the time of arrival of a particle. Indeed, 
if it occurred, the particle would never arrive, since the 
continuous measurement would project the particle on 
the particle-has-not-arrived state.
\bibitem{EXPLACOOLING} In Ref.~\cite{GSI}, the decay rate is measured only after
the first few seconds. The first few seconds are needed for cooling.
\bibitem{EXPLATOF} Equations~(\ref{timeofflig}) and~(\ref{timeoffligre}) 
also show that there 
exist physically relevant quantities, such as the time of flight, that are 
not associated with a single eigenvalue of a time operator but rather
with a mean value. See also
review~\cite{RECAMI}, where tunneling times are defined by way of mean
values.
\end{thebibliography}
\end{document}